# Mapping the Sahelian Space

Olivier J. Walther[1] and Denis Retaillé[2]

**Abstract**

This chapter examines the geographical meaning of the Sahel, its fluid boundaries, and its spatial dynamics. Unlike other approaches that define the Sahel as a bioclimatic zone or as an ungoverned area, it shows that the Sahel is primarily a space of circulation in which uncertainty has historically been overcome by mobility. The first part of the paper discusses how pre-colonial empires relied on a network of markets and cities that facilitated trade and social relationships across the region and beyond. The second part explores changing regional mobility patterns precipitated by colonial powers and the new approach they developed to control networks and flows. The third part discusses the contradiction between the mobile strategies adopted by local herders, farmers and traders in the Sahel and the territorial development initiatives of modern states and international donors. Particular attention is paid in the last section to how the Sahel was progressively redefined through a security lens.

**Acknowledgements**

The authors thank the OECD for their permission to use some of the maps of the *Atlas of the Sahara-Sahel*, and Al Howard for his useful comments on an earlier version of the chapter.

**Indivisible Sahel and Sahara**

Located far away from the seaports of the Gulf of Guinea and separated from North Africa by the world's largest desert, the Sahel has often been envisioned as an arid strip of land in which sedentary and nomadic peoples clash periodically. The global humanitarian response that followed the famines of the 1970s and 1980s temporarily brought the Sahel into international headlines and initiated a durable investment in food security and new forms of nongovernmentability (Mann 2014) but the region remained a peripheral area of interest in the scholarly community. The deterioration of the security situation that started with the emergence of armed groups affiliated with Al Qaeda and the Islamic State in the 2000s

---

[1] Visiting Associate Professor, University of Florida, owalther@ufl.edu
[2] Professor, ADESS Laboratory, University of Bordeaux and CNRS, denis.retaille@cnrs.fr

Walther and Retaillé

marked a radical change in the way the Sahel was perceived by scholars and policy-makers alike. After such a long period of obscurity, the Sahel – often confused with the Sahara – is now considered the center of a toxic combination of violent extremist organizations, secessionist rebels, and criminals, who exploit the weakness of the states and their porous borders (Walther and Miles 2018).

This chapter examines the geographical meaning of the Sahel, its fluid boundaries, and its spatial dynamics. Unlike other approaches that define the Sahel as a bioclimatic zone or as an ungoverned area, the chapter shows that the Sahel is primarily a space of circulation in which uncertainty has historically been overcome by mobility. By focusing on the interactions that the region and its people have with the broader North and West Africa, the chapter adopts a relational view on the Sahel that aim to avoid the "territorial trap" (Agnew 1994) that results either from considering the Sahel as a fixed territorial unit with strict boundaries or dividing it into mutually exclusive states. On the contrary, the chapter stresses how the Sahel constitutes a crossroads that has historically been linked to the rest of the African continent, particularly the Sahara, with whom it forms an indivisible pair, divorced only by colonial science.

The first part of the paper discusses how pre-colonial empires relied on a network of markets and cities that facilitated trade and social relationships across the region and beyond. The second part explores changing regional mobility patterns precipitated by colonial powers and the new approach they developed to control networks and flows. It shows that the metaphorical origin of the Sahel – which stems from the Arabic *sāḥil*, meaning "shore" – was ignored by the geographers, ethnographers, botanists, and explorers who attempted to define the limits of the Sahel by relying on natural criteria. The third part discusses the postcolonial period and the current contradiction between the mobile strategies adopted by local herders, farmers and traders in the Sahel and the territorial development initiatives of modern states and international donors. Particular attention is paid in the last section to how the Sahel was progressively redefined through a security lens.

**Pre-colonial Empires: The Sahel-Sahara as a Network**

The pre-colonial period of the Sahel-Sahara was characterized by a succession of empires for whom political power and commercial wealth was not a product of territorial domination but of control of the major routes that permitted the trade of gold, salt, slaves, dates, textiles, and wheat across the region (Austen 2010). The boundaries of these precolonial states were very different from those of modern Sahelo-Sahara states: fluid and flexible, they separated the main center of religious and political power from its vast peripheral areas, where slaves were raided.

*Precolonial empires*

Four great empires succeeded one another in time and space and controlled travel between North and West Africa between the end of the 9th century and colonization: the Ghana, the Mali, the Songhay, and the Kanem-Bornu (Map 1).

Map 1. Precolonial routes, cities and empires

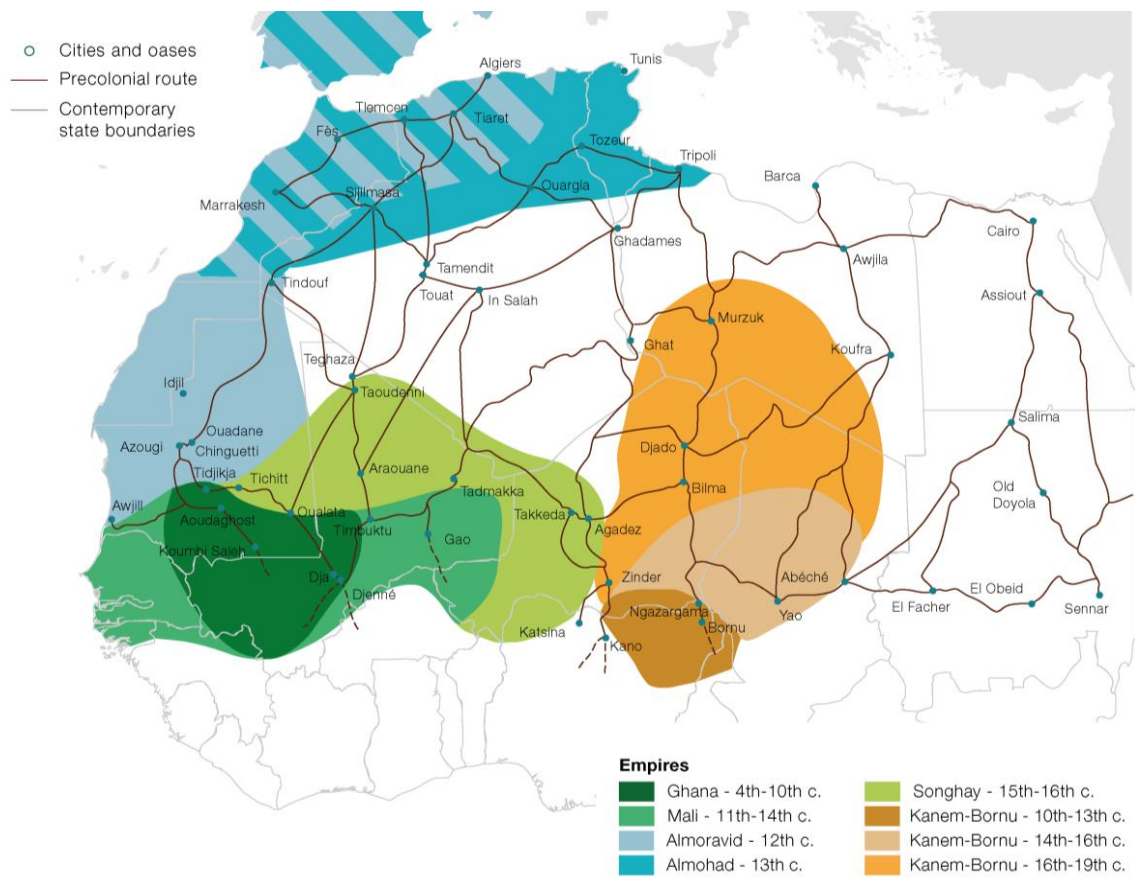

Source: Adapted from OECD/SWAC (2014) with permission.

The Ghana Empire was the first real trans-Saharan empire (Davidson 2014). It controlled the western routes from Koumbi Saleh to Sijilmasa before the nomadic Arabic population settled the area. It was replaced by the Almoravid Empire in the 12th century, which conquered Southern Spain and the western half of Maghreb. The transformation by the Almoravids of what had been a empire to a more traditional territorial form precipitated internal schisms that ultimately undermined their political power and resulted in the fragmentation of the empire into provinces.

In the 9th century, the Mali Empire succeeded the Ghana Empire without actually conquering it. The emergence of the Mali Empire occurred when the dominant Mali dynasty, the Keita, seized control of the trade routes in the region and shifted them from the Senegal River basin to the Niger River basin. As in the west, the population along this central route was made up primarily of camel drivers, ancestors of the ancient Sanhadja and Garamantes, as well as other Berbers fleeing Arab conquest. These groups took a long time to knit into a cultural community and never attained a significant degree of political unity.

Centered on Gao, the Songhay Empire (15-16th c.) was the last of the empires based on the Senegal and Niger Rivers (Abitbol 1999). It was the most extensive of the Sahel-Saharan empires and controlled the same central trans-Saharan route as the Mali. The Songhay Empire was dominated by the Touré, known as Askia, whom at their peak in the early 16th century ensured the stability and permanence of movement within the Sahara at a time when the Maghreb, which collapsed after the Almohad episode, was still unstable. The Moroccan conquest that brought an end to the Songhay in 1591 radically changed the centrality of the trade network to the profit of the North. It revived within a single entity the large central Timbuktu-Sijilmasa route, passing through the salt mines of Taghaza (later and further to the south, Taoudenni) via the Tanezrouft desert.

A thousand kilometers east, the most enduring of the empires, the Kanem, and subsequently the Kanem-Bornu (10-19th c.), was centered on Lake Chad but extended as far as the current route from Chad to Sudan (Barkindo 1999). The Tubu and Kanuri who built the empire ensured the permanence of movement within the Sahara by controlling the route to Fezzan in present-day southern Libya, the main Saharan crossroad between Lake Chad and the Mediterranean Sea. Despite being less well known than other pre-colonial routes, this route has been one of the most durable axes of commerce of the region.

This succession of Sahelo-Saharan empires demonstrates how history often repeats itself. First, a powerful dynasty like the Keita of Mali or the Touré of Songhay or an organization like the Almoravids managed to gain control over commerce through alliances. The accumulation of commercial wealth then permitted them to establish political power. This situation was, however, precarious as the social groups excluded from power perpetually contested the ruling authority until power shifted to another dynasty. This succession of dynasties was first described by Ibn Khaldun (1378/1969), who noted that the "sedentarization" of power over movement was accompanied by a softening of nomadic warrior spirit and pride. The experience of the Sahelo-Saharan empires shows that attempts to transform routes into conduits of power radiating from central places results in the leakage of power to groups that are unencumbered by the challenge of controlling sparsely populated territory.

*Routes, population centers, and mobility*

Precolonial empires were not dedicated solely to commerce. Each empire was anchored in a densely-populated area in the Sahelian zone, from which the trans-Saharan routes to the north and southern routes to the savannah and Gulf of Guinea radiated. The Ghana Empire was centered in the west of modern Mali and to the south of Mauritania, the Mali and Songhay empires were governed from the Niger River valley, and the Kanem-Bornu Empire from the Lake Chad region. These population centers were linked to the large nomadic areas of the Sahara: the Senegal River population basin offered access to the Moorish nomadic territories, the Niger River valley was connected to Tuareg grazing lands, the Hausa settlements offered access to the Azawagh of northern Niger, and the Tubu settlements to the Djado.

Over the centuries, particular patterns of mobility developed within each empire and at their fluid boundaries. Sahelian populations developed trade networks linking the principal salt, textile, kola nuts, and natron producing areas with regional consumer markets. In addition to the Hausa merchant diaspora organized around kola (Lovejoy 1980) several other large networks operated between the basins of the Volta and the Niger Rivers: the Dyula from the Mali Empire, the Yarse from the Mossi kingdom, and the Wangara who were associated with both the Ghana and Mali empires.

Farmers and pastoralists also engaged in seasonal exploitation of the rich grazing land and cultivation areas. Their patterns of settlement and periodic movements were finely tuned to the agro-ecological areas they connected and knew exceptionally well. In the Inner Niger Delta, for example, three systems of production were intertwined. Depending on the season, fishermen, farmers and pastoralists invested the flooded areas, the fertile lands and the grassy meadows of the delta each in its turn. Formalized in the 19$^{th}$ century with the founding of the

Walther and Retaillé

theocratic empire of Macina, this socio-political organization shunned the intensive cultivation of areas that, elsewhere in the world, would have attracted a significant and permanent rural population (Gallais 1967). In addition, agricultural expansion strongly contributed to define property rights and the structure of relationships between social groups in the region. In many regions, such as the Dendi, Northern Ghana and Southwestern Burkina Faso, conquerors who supplanted local chiefs became responsible for political authority, while the "sons of the soil" retained their religious power over animist cults, land administration, and natural resources (Walther 2012, Lentz 2013).

A number of urban centers where food stuff and cereals were purchased by the nomads in exchange for salt and animals punctuated these spaces. Cities located at the crossroads of caravan routes functioned as strategic places for the circulation of goods and people: Bilma and Iferouane in Niger, Chinguetti, Tichit, Oualata and Atar in Mauritania, Tamanrasset in Algeria, Gao in Mali (Lydon 2009). The alignment of these centers in the southern Sahara, just as in the north, constituted a network of places that served as transit centers between the desert and the savannah (Retaillé, 1995, OECD/SWAC 2014). This particular spatial structure in which trans-Saharan routes were associated with marketplace remained remarkably stable until the dawn of the colonial period, despite the temporal and spatial succession of empires.

**Exploration and Colonization: Dividing the Sahel-Sahara**
Nineteenth century European explorers, attracted by the accounts of ancient and Arab geographers, followed the routes used by the nomadic peoples of the Sahara. They documented the names of the stopping places they discovered, described landscapes, took climate measurements, and gathered rock samples. Gradually, a map including topographical relief and the location of watering holes crucial to long crossings began to fill in the blank space that represented the region in European atlases. The Arabic word *Sahel* made its entry into the vocabulary of geography. In the north, as well as the south, the search for the limits of the Sahel and the Sahara was undertaken for the twin objectives of finding *routes* across the immense region and the *boundaries* that would enable its division. It was both a scientific and political exercise.

*Mapping the routes and boundaries of the Sahara*

The routes traveled by all explorers through the 19th and into the 20th centuries followed those traced, in previous centuries, by the pre-colonial empires and nomadic caravaners. Later on, the colonization of this vast territory, primarily by the French, also involved trans-Saharan railroad and road projects based on the dream of linking the two colonized areas of Maghreb and West Africa. Due to internal rivalries between French colonies, these ambitious projects never materialized and both shores of the Sahara progressively became completely disconnected from each other (OECD/SWAC 2014). At the same time, colonial era geographers attempted to identify the borders of the vast desert in the north and south and establish sensible subregions based on human and climatic characteristics, like rainfall and sedentary and nomadic "ways of life".

In their search for the boundaries of and passageways through the great desert, explorers and conquerors considered the desert a barrier between two great hydrographic systems – one oriented towards the Mediterranean and the other to the Gulf of Guinea. The seasonal rivers that penetrated the desert made up a vast basin that never flowed into the sea, with the exception of the Nile River. This vocabulary of geographical positivism was inspired by the work of Philippe Buache (1752), the geographer of the French king Louis XVI, who used hydrographic basins to delimit regions. The boundaries of specific regions were, therefore, defined by the network formed by the principal rivers and their tributaries from their sources to the sea.

Features of the terrain played an equally important role in this approach to understanding space, particularly for French explorers and colonizers for whom the Massif Central was considered both the geological (Dufrénoy and Elie de Beaumont 1841) and historical center of France (Michelet 1886). The presence of the Hoggar and Tibesti Mountains in the center of the Sahara was reminiscent of the continental upland. This similarity was noted by the French explorer Conrad Killian (1925: 11), who in his description of Hoggar noted that "to reach the Saharan Massif Central you must first traverse a whole country of vast plains". The European geomorphological model served as a point of references for interpreting the geography of the Sahara and the identification of "massifs centraux" gained currency as a method for constructing geographical frameworks alongside rivers and other boundaries and barriers. Later on, the Sahara was conceptualized as a "roof" with two pitched sides, one to the Mediterranean, and one to the "Sudan" in the south, and an edge oriented NE-SW. From the Atlantic Ocean to the Nile, geographers also identified three large empty areas separating nomadic populations: the Tanezrouft between the Moore and the Tuareg, the Ténéré between the Tuareg and the Tubu, and the Libyan Desert in the Far East (Monod 1968).

This representation of the Sahara as a roof punctuated by central mountains and scattered human settlements was frequently used to delimit nomadic "countries" (*pays* in French). The *pays* Tuareg, an object of considerable attention for the French, was for example defined as a region with "fairly precise boundaries in the heart of which [the Hoggar, the Aïr and the Adrar des Ifoghas] form an armature" (Bernus 1991: 120). In addition to forming the high country of the nomads, the central mountains of the Sahara were considered as a starting point from which to fix the limits of sedentary society, where movement could be controlled and territory could be conceivably divided for agricultural exploitation.

*The invention of a frontier between sedentary and nomadic populations*
The delimitation of a frontier between sedentary and nomadic populations was the object of fierce debate during the colonial period. North of the Sahara, this debate engaged the geographers at the University of Algiers who were, more or less, devotees of the French School of regional geography espoused by Paul Vidal de la Blache. The "Algerians" led by Emile-Félix Gautier (1923) supported meticulous field survey. They eventually prevailed over the "Parisians" represented by Augustin Bernard and Napoléon Lacroix (1906), whose main preoccupation was to draw a line between the fertile production lands extending along Algeria's coast and the more rebellious Sahara. The desert boundaries were therefore established at the first line of oases visited seasonally by nomadic tribes that punctuate the foothills of the Atlas Mountains. The limit draws on the historical boundary set by the Regency of Algiers, also known as Ottoman Algeria (c. 1517-1830), that prevailed until colonization. Other administrative regions were created in the Sahara: between 1902 and 1952, French Algeria constituted the Southern Territories, a military area patrolled by companies of camel corps against nomad incursions.

In the south, geographers also sought the boundary of the Sahel and the Sahara, this time on a purely bioclimatic basis and with the implicit idea that rainfall conditioned sedentary or nomadic "ways of life". Using maps to record the volume of rain that fell each year made it possible to draw lines connecting locations that received equal amounts of rainfall, known as isohyets. Because the volume of water that falls on the surface of the earth varies a great deal over time, particularly for smaller volumes, rainfall was therefore expressed as an average annual normal. In this view, the 200 mm annual precipitation isohyet was defined as the cut-off for aridity, below which farming and herding were no longer possible. Regions receiving less than 200 mm of rain per year were considered Saharan while those receiving between 200 and 600 mm were considered Sahelian. Further south, the Sudanian domain was characterized by precipitation between 600 and 1300 mm, and the Guinean domain with precipitation greater than 1300 mm (OECD/SWAC 2014).

This zonal distribution is mainly due to the seasonal move of the Intertropical Convergence Zone (ITCZ), where northeasterly winds from the Sahara, called *harmattan* in the Sahel, and southwesterly winds from the tropical Atlantic Ocean converge (Lélé and Lamb 2010). The boundary between the dry and hot air of the continental trade winds and the warm, humid air of the southwesterly winds is roughly aligned east-west. Its movements condition the intensity of rainfall in the Sahel: when the trade winds move northward, the convergence zone moves in this direction until reaching its maximum position, approximately 20°N in the month of August. From April to October, West Africa experiences a rainy season, when the southern winds prevail and bring moist air to the interior of the continent. Starting in November, the zone of convergence begins its descent towards the equator. The Sahel then experiences a dry season, characterized by the preponderance of the Saharan winds, until the following April.

Because the relative strength of the northern and southern trade winds greatly varies from one year to the next, any effort to use isohyets to define the boundaries of the Sahel proved challenging. As a transition zone between the Sahara and the savannah, the Sahel is the region where the coefficient of variation for annual precipitation is the highest in West Africa (Map 2). Rainfall anomalies are the norm rather than the exception and wet seasons can be followed by droughts in a rather unpredictable way (OECD/SWAC and Met Office 2010). The 1968–2000 average normal, for example, shifted the isohyet lines about one hundred kilometers to the south, strongly suggesting that the desert was gaining ground. Data recorded in recent years suggest that moisture may be returning. Due to global climate change, however, the projections for Sahel rainfall are less robust than 20$^{th}$ century records, with a probable increase of extreme meteorological events such as massive rainfalls followed by periods of relative droughts (Panthou et al. 2014).

Map 2. Variation for annual precipitation in West Africa

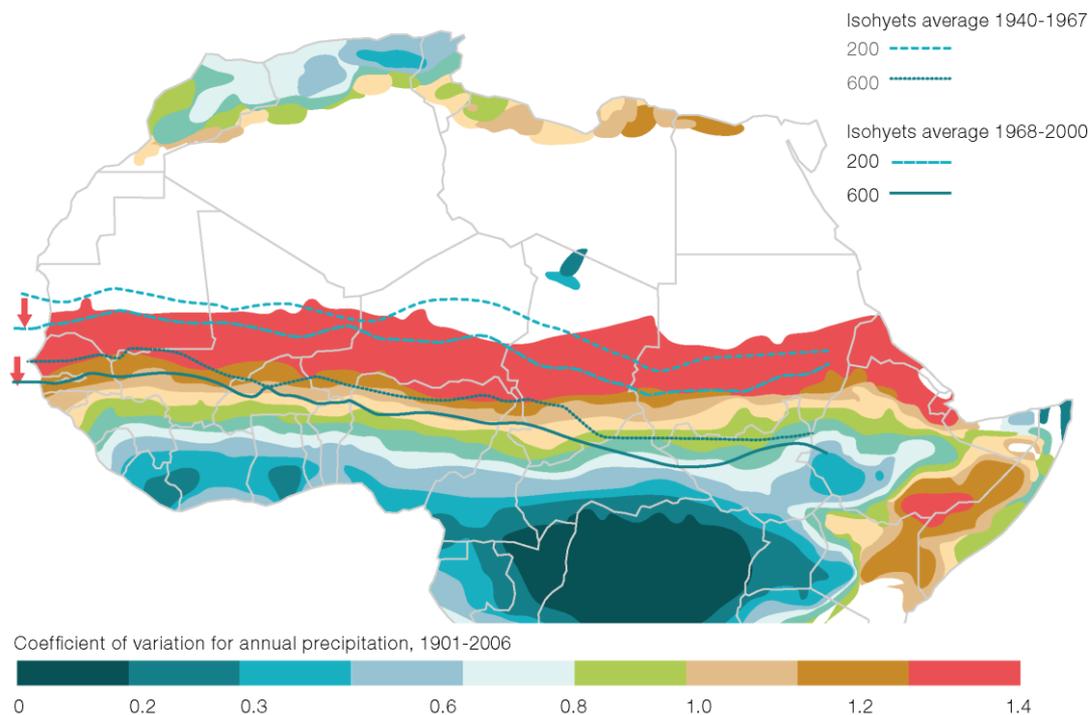

Source: Adapted from OECD/SWAC (2014) with permission. Coefficient of variation calculated by: UK Met Office Hadley Centre; University of East Anglia Climate Research Unit (CRU) and defined as standard deviation normalized by the mean (only regions with annual precipitation > 10 mm/year).

*The Sahel separated from the Sahara*

By casting settlement as the opposite of nomadism, geographers of the colonial era precipitated an ethnicization of ways of life and spatial practices. The nomadic existence of the herders was juxtaposed with the investment in land of the farmers and a civilizational fault line was drawn. In addition to being considered more productive than the nomads, sedentary populations were considered more receptive to the French "civilizing mission" (Benjaminsen and Berge 2004a). The settlement policies pursued by the French in the Sahel were in part an effort to control populations they considered too independent with respect to colonial power, and partly an attempt to promote a development model centered on agriculture and mineral resources. The nomads were subject to rules defining the administrative region to which they would be attached for the purpose of paying taxes, immunizing their herds, and educating their children and to circumscribe the limit of their nomadic "territories" (Clauzel 1992). Cartographic boundaries and strict climatic zones masked the circulation that was characteristic of the region and offered it the flexibility to survive periods of political and

10
Walther and Retaillé

climatic instability. Thus, the Sahel was separated from the Sahara as if they had nothing at all to do with one another.

This view contradicts the spatial dynamics observed in the Sahel and Sahara during the colonial era. Because of bioclimatic fluctuations, herders and farmers were intertwined and often occupied the same areas depending on the seasons (Retaillé 1989). Both tended to disperse from one another during good years or seasons: the farmers expanded their agricultural frontiers while the herders spread out to abundant pasture lands. Convergence occurred during dry seasons or periods of drought: farmers left peripheral hamlets to regroup in villages, while the nomads sold their herds and head towards urban centers. Mobility also varied considerably among sedentary and nomadic populations. For the Moore, Touareg, Tubu, and Fulani, sedentary lifestyles were actually very common (Bernus 1993). While the poor were devoted to agricultural work in the oases, rich and noble Saharan were not compelled to move incessantly in search of pasture for their cattle. Thus, the Sultan of Agades has always lived in town, just as the wealthy Tuareg noble families had limited nomadic patterns and delegated the breeding work to people of lower social status. Sedentarization was thus done either from the bottom, due to the impoverishment of nomadic groups that the countryside could no longer feed, or from the top, through the enrichment and the transfer of tasks to serving groups.

Among many sedentary societies, mobility was just as important. For the Soninke, Marka, Hausa and Kanembu, being mobile was deeply ingrained, permitting traders and migrants to participate in commercial activities and inter-regional flows of migration (Miles 1994, Manchuelle 1997). Traders and migrants demonstrated a great adaptability to the 'new spatial order' brought by colonization (Howard and Shain 2005). After the French and British drew borders between their West African colonies, trade networks were re-oriented to supply newly established colonial cities, use new railway and roads, and avoid customs authorities. The creation of border posts between the Colony of Niger and Nigeria in the mid-1910s, for example, led Nigerian traders to relocate south of the border, where better business opportunities were available, until the removal of border controls in the 1950s encouraged the massive development of cross-border trade (Nugent and Asiwaju 1996).

11
Walther and Retaillé

**Sahelo-Sahara states**

The bioclimatic definition of the Sahel of colonial inspiration was hardly contested by post-colonial elites, who enthusiastically embraced the distinction between nomadic and sedentary populations and the rigid demarcation between the Sahel and the Sahara. A similar view was shared by foreign donors and international organizations, for which development necessarily implied a stronger investment in land and other localized resources.

*The end of the nomadic world*

Many of the new Sahelo-Saharan states that emerged during the 1960s promoted production to the detriment of pastoral migration and pursued settlement policies against the nomads. Such states policies were designed to make the patterns of settlement and social life of nomadic people legible to the state (Scott 1998). The idea was that the state could more efficiently provide public services, prevent rebellions, and promote intensive forms of agriculture and cattle breeding if population was assembled into permanent settlements. Drawing on the zonal model of colonial inspiration, states adopted laws that limited the agricultural expansion in the north and constrained pastoral movements to the south.

In the Republic of Niger, for example, law 61-05 of the Code Rural, which builds on a similar colonial order of 1954, defined a limit to the north of which "all rainfed crops and installation of farmers' groups are prohibited" (République du Niger 2013: 41). Similar decrees adopted in 1961 defined how water from pumping stations and grazing lands had to be used by herders (République du Niger 1961). In Mali, Modibo Keita's government considered the nomadic lifestyle as unproductive and as an obstacle to modernization and launched a series of ultimately ineffective measures to settle the Arab, Fulani, and Tuareg populations. In the Inner Niger Delta, these policies attempted to transform the nomads' seasonal grazing lands into irrigated rice fields (Benjamin and Berge 2004b). In Nigeria, federal policies also aimed at intensifying production by adopting a ranch model and settling pastoralists, with limited success considering the cross-border mobility of many nomadic populations (Blench 1996).

By the mid-1970s, social scientists started to recognize that while tracking annual and periodic fluctuations of the isohyets had permitted them to map the agricultural and natural resources of the Sahel it had also obscured social dynamics, particularly in time of crises such as the great droughts (Monod 1975, Gallais 1984). During this period, the rainfall deficit in the Sahel caused the loss of a large part of the herds deprived of water and pasture. Struggling with their means of subsistence and exchange, the Saharan nomads were forced to settle in the southern Sahelian zone, in Algeria, or in Libya. Migratory movements to cities, which only concerned certain servile categories before the great droughts, such as tributaries, or some marginal elements such as divorced women and adolescents in school, suddenly extended to large parts of the nomadic societies, leading to the weakening of cultural and economic life in the rural world.

The droughts of the 1970s also marked the beginning of large-scale international interventions, encouraged by the Western media. At that time, humanitarian and international organizations regarded sedentarization as inevitable and financed agricultural projects dedicated to convert ruined nomads into agriculturalists. These initiatives to permanently settle mobile people rarely succeeded. Their economic and environmental failure was often explained by the fact that they paid little attention to the local knowledge, social practices and spatial representations of the pastoralists and cultivators of the Sahara-Sahel. Herders for whom cattle was a sign of wealth and social prestige proved reluctant to adopt a new sedentary lifestyle. When the aid agencies or the state managed to settle nomadic groups, it was generally only until they had reconstituted their flocks, or on other occasions, only the former slaves settled in the villages.

*Mobility rediscovered*

The last 30 years have provided an opportunity for geographers and other social scientists to rediscover original forms of mobility that had been obscured by the humanitarian crises of the 1970s and 1980s. This rediscovery has led to a new geography of the Sahel and the Sahara, more attentive to the mobility and adaptability of local societies to climatic variations. The research carried out in the region has stressed the conceptual limitations of the zonal division of West Africa and the need to consider it as a single circulation area that transcends ecological zones. Anthropologists have highlighted the sheer adaptability of pastoralists, who favor flexible and opportunistic responses to climate and political uncertainties (de Bruijn et al., 2001, de Bruijn 2007, Boesen and Marfaings 2014), as well as the smuggling and semi-official trade networks that crisscross the Sahara (Scheele 2012, McDougall and Scheele 2012). Other social scientists have documented the mobility patterns of the Sahelo-Saharan populations (Gagnol 2012), with a particular focus on cross-border networks (Choplin and Lombard 2014), Trans-Saharan migration (Marfaings and Wippel 2004, Bensaad 2009) and trade (Grégoire and Labazée 1993, Grégoire 2002).

Drawing on an analysis of the traders' spatial strategies and the historical dynamics of the pre-colonial commercial system, historians have developed what is known today as the 'spatial factor' approach to African history (Howard 2010, Howard and Skinner 1984, Howard and Shain 2005). The spatial factor approach highlights how traders were quick to explore new markets and adjust to changing economic opportunities and market conditions due to the flexibility of their social networks. It shares many similarities with the 'mobile space' approach initially developed by geographers to analyze the effects of the droughts on the livelihoods of farmers and herders in the Sahel-Sahara (Retaillé 1989) and, later, progressively formalized to reflect the primacy of mobility in the spatial organization of the region (Retaillé and Walther 2011, 2013, 2014, Retaillé 2013). Both approaches emphasize the role of Sahelo-Saharan markets, viewed as "nodes of transregional or transnational trade networks and places in production territories" (Walther et al. 2015: 347).

**Security and Sahel strategies**

In the early 2000s, the deterioration of the security situation in the region dramatically increased the academic, military and policy interest for the mobility strategies of the Sahelo-Saharan populations. Scholars and analysts quickly adopted the terms Sahel-Sahara or Sahara-Sahel to describe the interdependencies between the two regions (Lacher 2012, Zoubir 2012, IPI 2013, OECD/SWAC 2014, Harmon 2014, Reeve and Pelter 2014, Boukhars 2015). Foreign powers intervening in the Sahel-Sahara also adopted a more transnational approach to the region, while African and international organizations drafted "Sahel strategies" that mainly conceived of the region as a collection of states.

*Countering transnational threats in the Sahel-Sahara*

The geographical expansion and opportunistic relocation of many violent extremist organizations, such as Al Qaeda in the Islamic Maghreb (AQIM) and Boko Haram, have led foreign powers to challenge the hard edge between Sahel and Sahara that was so prevalent during the colonial and postcolonial periods. Both the United States and France have adopted counterterrorism strategies and military tactics that focus on the fundamentally transnational nature of the threats in the region.

The United States recognized the risk inherent in the rise of Sahara-Sahel terrorism shortly after the 9/11 attacks. In 2002, the United States Department of State implemented the Pan

Sahel Initiative (PSI) to support border control capabilities, control illicit trade and enhance regional security between Mali, Mauritania, Niger and Chad (Map 3) (Ellis 2004). After the Salafist Group for Preaching and Combat (GSPC), later renamed AQIM, conducted an impressive hostage-taking in the Algerian desert in 2003, the PSI was turned into a counter-terrorism program for the states that had at least part of their territory located in the Sahel region and were threatened by the movements in the Sahara. PSI was replaced by the Trans-Sahara Counterterrorism Initiative (TSCTI), a much larger program that involved, in addition, the Maghreb states, as well as Burkina Faso, Senegal and Nigeria, and was incorporated in the United States Africa Command (AFRICOM) established in 2007 (Francis 2010, Harmon 2014).

The regional dimension of the terrorism threat is also at the heart of the initiatives developed by France after Operation Serval was able to regain control of northern Mali in 2013 and paved the way for the United Nations Multidimensional Integrated Stabilization Mission in Mali (MINUSMA) and the European Union Training Mission in Mali (EUTM). Operation Barkhane, which succeeded Operation Serval in August 2014, explicitly addresses the regional and cross-border dimension of terrorism activities in the Sahel-Sahara. As noted by its former Commander, General Jean-Pierre Palasset, Barkhane differs from previous military engagements that targeted "one country, one crisis, and one theatre of operations" (Ministère de la Défense 2014).

The 4000-strong contingent of French forces relies on three ports in the Gulf of Guinea (Dakar, Abidjan, Douala) for their supply, two airports in the Sahel (N'Djamena and Niamey) where fighter jets and drones are stationed, and several Saharan outposts such as Madama in the trinational area between Niger, Chad and Libya, from which cross-border trafficking routes and terrorist networks can be disrupted (Map 3). A truly regional operation, Barkhane is nonetheless facing serious logistical issues related to the vastness of the area covered by the operation – nine times larger than metropolitan France – as well as the difficult climatic conditions in the Sahara which impose a heavy toll on the equipment and troops.

Map 3. U.S. and French military initiatives and violent political events, 1997-2015

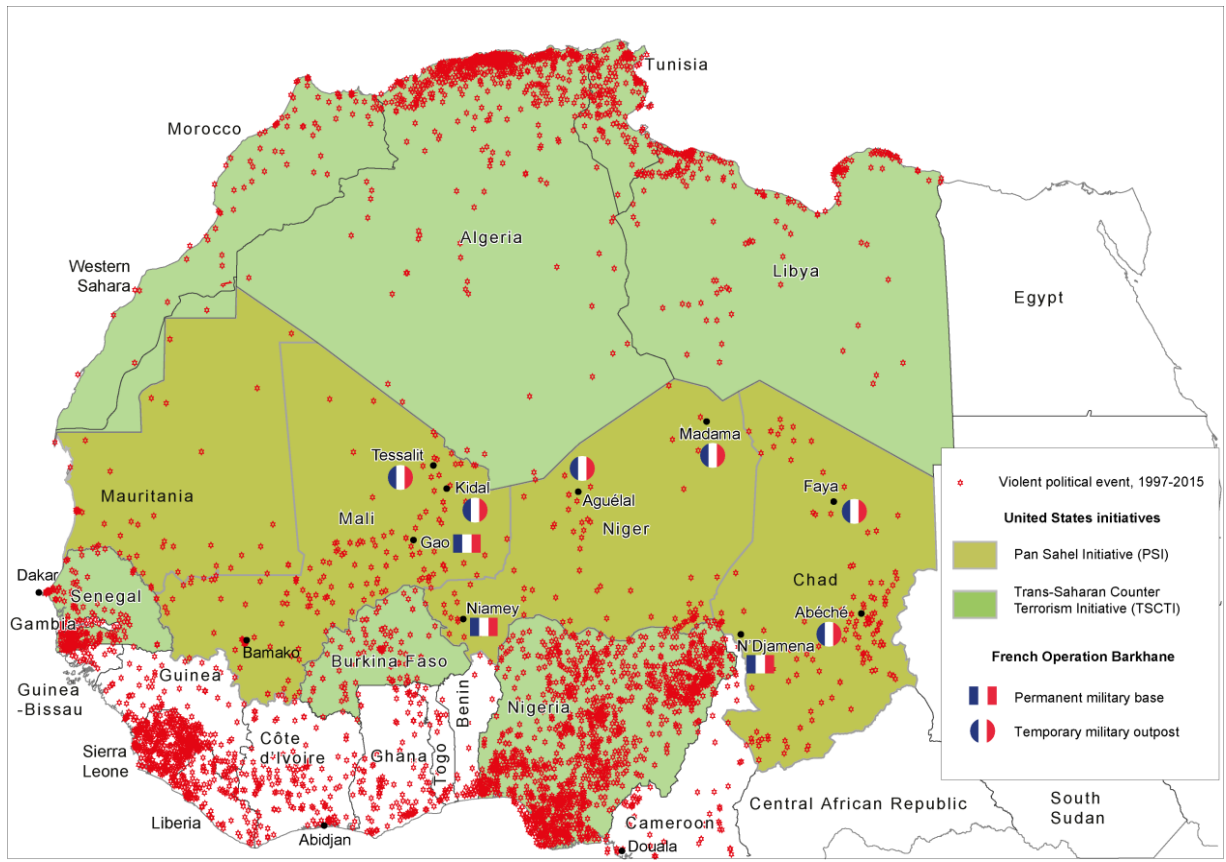

Source: Armed Conflict Location and Event Data project, US Department of State; French General Staff andf Ministry of Defense. Cartography: authors.

*The Sahel as a collection of countries*

From a policy perspective, the degradation of the security situation has led an increasing number of states and intergovernmental organizations to recognize that the problems faced in the Sahel can only be understood and solved with a regional approach. The European Union (2011: 1), for example, argues that "many of the challenges impact on neighboring countries, including Algeria, Libya, Morocco and even Nigeria, whose engagement is necessary to help resolve them". Similarly, one of the strategic goals of the United Nations Sahel Strategy (2013: 18) is to build and support "national and regional security mechanisms (that) are capable of addressing cross-border threats". The African Union (2014: 13) observes that the establishment of terrorist organizations and transnational criminals has "worsened the climate of insecurity and now constitutes the major threat in the region". These threats, the African Union argues, have "a regional and international dimension aggravated by the weakness of border control exercised by the States of the Sahel-Saharan sub-region and their lack of operational and strategic capabilities".

In order to address these issues, several 'Sahel' strategies have been initiated in the region. These include the Joint Operational Army Staffs Committee (CEMOC), an Algerian initiative created in 2010 to coordinate the fight against terrorism and criminality, the African Union's Joint Fusion and Liaison Unit put in place by the militaries of eight North and West African countries in 2010 to share information and coordinate activities against terrorism, weapons smuggling and narcotics, and the Nouakchott Process launched in 2013 with the aim of reinforcing security cooperation in the wider region. The European Union's Strategy for Security and Development in the Sahel (2011) and the United Nations Integrated Strategy for the Sahel (2013) are two other examples of initiatives that combine governance, security and development at the regional level. Other strategies include the Strategy for the Sahel elaborated by ECOWAS, UEMOA and CILSS, the Strategy for the Sahel Region of the African Union and the Sahel G5, adopted in 2014 (Map 4).

There is a little consensus among states and intergovernmental organizations as to which region the Sahel strategies should target (Helly et al. 2015). Geographically, their only apparent common denominator is that they should build on a number of *countries* in which security issues have been identified. By doing so, Sahel strategies have contributed to challenge both the definition of the Sahel as a bioclimatic zone and crossroads. The African Union (2014: 2), for example, focuses on "all the countries located on the Sahelian strip separating North and Sub-Saharan Africa". Outside this area, it notes that there are countries "who deserve, all the same, special attention, including Cote d'Ivoire, Guinea, Guinea-Bissau, Nigeria and Senegal". International organizations such as the European Union, the United Nations and the World Bank have identified a number of "Sahelian countries" that include Mauritania, Mali, Burkina Faso, Niger and Chad. With the exception of Burkina-Faso, all the countries comprise vast expanses of land located in the Sahara.

Map 4. Sahel strategies

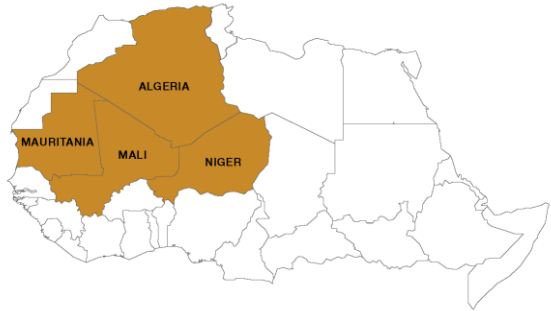
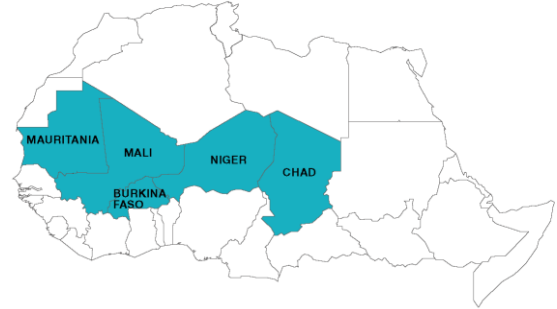
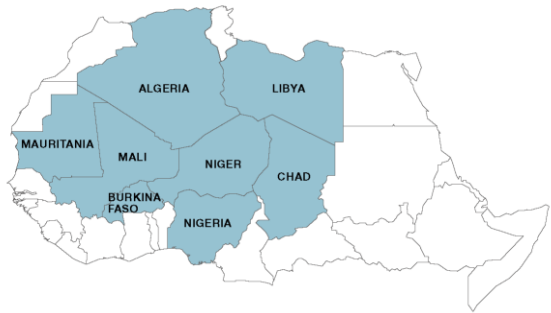
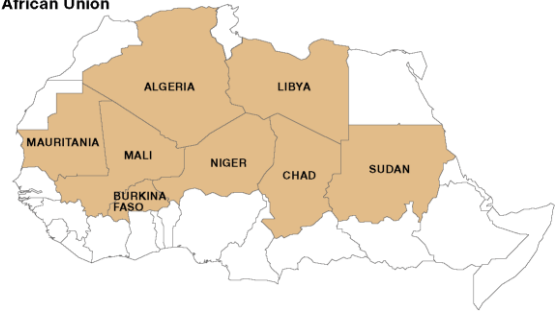
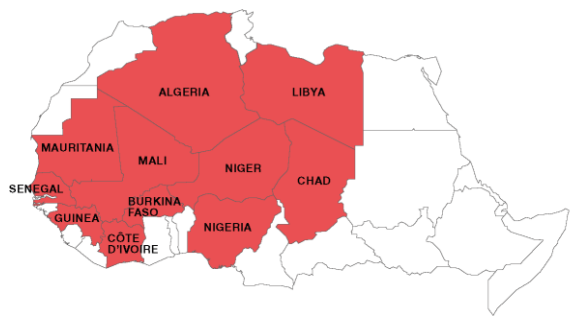

Source: Adapted from OECD/SWAC (2014) with permission.

The choice of which countries should be included in the Sahel strategies more often reflects regional rivalries and state agreements, both within Africa, and with Western countries, than real functional interdependences within the region. For example, Morocco is excluded from most Sahel strategies despite being a key Western ally in North Africa, due to its rivalry with Algeria and long exclusion from the African Union. An Algerian creation headquartered in Tamanrasset, the CEMOC is limited to Algeria and its southern neighbors, without including Libya, which has been one of the epicenters of regional political instability since the fall of Col. Gaddafi in 2011. Senegal, a purely Sahelian country, is only part of the Nouakchott Process.

While military and policy initiatives differ in their own definition of the Sahel, they share much in common in terms of how they conceive the cultural and social relationships that develop within the region. Both have largely disregarded the classical distinction between sedentary and nomadic populations elaborated by colonial science, to the profit of another binary division between radicalized individuals and the rest of society. Particular attention is paid to the drivers of violent extremism (USAID 2011), including social marginalization, government repression and human rights violations, endemic corruption and elite impunity.

**Conclusion**

Since the end of the 19th century, geographers have attributed the Sahel with a double geographical identity. Based on variations in mean annual rainfall, the Sahel was first defined as an intermediate *zone* between the desert and the savanna that extended from the Atlantic to the Red Sea. The Sahel was also defined by colonial geographers as a *front* between sedentary and nomadic populations. Colonial science and policies relied on these two attributes to progressively map the location of the natural, legal and cultural boundaries of the Sahel. This interpretation of the Sahel has not been questioned by the post-colonial policies supported by international organizations, which have been focused on settling nomads, limiting their rangelands and enhancing the agricultural potential of areas where rainfed and irrigated agriculture was possible.

The catastrophic droughts of the second half of the 20th century have gradually called into question the idea that the Sahel was a territory that could be divided, appropriated and developed. Studies carried out on the pastoral, agricultural and trade movements of the Sahelo-Saharan populations from the end of the 1980s show that the urgency of humanitarian crises had masked the fact that local populations did not necessarily share sedentary representations of space and often privileged movement over territorial fixity. What really mattered to them was the ability to circulate across the region and create a flexible social network that transcended national and climatic boundaries. It has been increasingly clear that the Sahel could be defined as a crossroads in which the spatial structure, made up of cities and markets, could adapt to climatic variations and political crises.

The development of terrorism and trafficking has made the search for rigid boundaries even more elusive, blurring the distinction between the local, the national and the global. Violent extremist groups in the region thus rely on certain pan-Islamic ideas, such as jihad, and on their affiliation to Al Qaeda and the Islamic State to express local or national grievances. Their attacks are seemingly unpredictable because they are based on a high degree of mobility and opportunistic use of national borders. Traffickers who smuggle migrants, weapons and drugs across the region also rely on transit centers and trans-Saharan roads that rapidly adapt to the actions of states. None of these networks is purely Sahelian.

The degradation of the security environment in the region and the proliferation of "Sahel" strategies devoted to addressing political instability have prompted further questions about what the Sahel really was. Military strategies implemented by the United States and France since the 2000s seem to suggest that the Sahel is the epicenter of a large theatre of operations that virtually knows no boundary. The fact that violent extremist organizations can expand or relocate to a neighboring country where military capabilities or political will are weaker means that the bioclimatic zones upon which the Sahel was defined for much of the 20$^{th}$ century have lost their importance. Policy makers in charge of designing Sahel strategies have also largely disregarded bioclimatic zones, to the profit of a new division by country. Such strategies stress the internal divisions of the Sahel-Sahara rather than its functional interdependencies, that transcend state boundaries, from the Gulf of Guinea to the Mediterranean Sea.

The near-impossibility to define the Sahel in a purely territorial manner has consequences for scholars and policy makers alike. If the Sahel and its people have no boundaries and can move relatively freely from Abidjan to Tripoli, new geographic tools should be developed to represent and model transnational patterns that are not, as with many existing maps, limited to fixed points, straight arrows, and states boundaries but can also include dynamic features such as social ties, alliances and conflicts, and spatial representations. Sahel strategies that currently built on mutually exclusive states should also move beyond territorial approach to regional security and more thoroughly address the roots, structure and resilience of the social and spatial networks that enhance political violence and trafficking in the region.

Walther and Retaillé